# Three-dimensional $x-y$ model with the Chern-Simons term

Norbert Schultka and Efstratios Manousakis

*Department of Physics and Center for Materials Research and Technology*
*Florida State University, Tallahassee, Florida 32306*

November 14, 1994

**Abstract**

We investigate the influence of the Chern-Simons term coupled to the three-dimensional $x-y$ model. This term endows vortices with an internal angular momentum and thus gives them arbitrary statistics. The Chern-Simons term for the $x-y$ model takes an integer value which can be written as a sum over all vortex lines of the product of the vortex charge and the winding number of the internal phase angle along that vortex line. We have used the Monte-Carlo method to study the three-dimensional $x-y$ model with the Chern-Simons term. Our findings suggest that this model belongs to the $x-y$ universality class with the critical temperature growing with increasing internal angular momentum.

## 1 Introduction

The $x-y$ model in two and three dimensions has been studied intensively (cf. Ref. e.g. [1]). It was found that vortex excitations play a crucial role in the phase transition. Kosterlitz and Thouless showed that the pairing of vortices in the $x-y$ model causes the phase transition in two dimensions [2]. In three dimensions the vortex strings seem responsible for the occurence of the phase transition [3], which has the character of an order-disorder phase transition. The remarkable property of the $x-y$ model is that the topological objects, vortices in two and vortex lines in three dimensions, which are responsible for the occurence of the phase transition, do not explicitly appear in the model. They are rather created dynamically and extended over long length scales. Thus we can consider the $x-y$ model as a microscopic field theory which contains quasiparticles which are extended objects over long length scales and interact through a potential created by the interaction of the microscopic degrees of freedom. In the two-dimensional $x-y$ model the quasiparticles are the vortices which interact through a logarithmic potential, while the microscopic degrees of freedom are elementary and are coupled by a nearest neighbor potential (cf. the next section). The three-dimensional $x-y$ model can then be viewed as describing two-dimensional vortices moving in euclidean time. By coupling the three-dimensional $x-y$ model to the Chern-Simons term we can give the vortices an internal angular momentum taking values between 0 and 1/2. Since the internal angular momentum of the particles is closely related to statistics, quasiparticles of various statistical character which interpolates between that of a boson and a fermion, are dynamically created. It is interesting to investigate how the thermodynamic properties of the system containing vortices changes when these vortices obey different statistics.

The effect of the Chern-Simons term of endowing extended topological objects with an internal angular momentum is not restricted to our model. In three dimensions the proper coupling of the Chern-Simons term to a conserved current introduces fractional statistics of the particles which are associated with this current [4, 5, 6, 7]. Since the statistics of these particles interpolates between Bose and Fermi statistics they are called anyons. Anyons are not only a theoretical concept, they occur in at least one physical system, namely, the quasiparticles in the fractional quantized Hall effect exhibit anyonic properties. The fractional quantized Hall effect can be described in the framework of a generalized Landau-Ginzburg theory, the so-called Chern-Simons-Landau-Ginzburg theory [8], where the electrons are minimally coupled to a statistics changing "electromagnetic" field whose action is the Chern-Simons term. Another example where the Chern-Simons term introduces fractional statistics of the quasiparticles is the nonlinear $O(3)$ model (or $CP^1$ model) in $2+1$ dimensions [6, 7]. In particular Wilczek and Zee showed in Ref. [7] that the solitons, the topological quasiparticles in this model, acquire a fractional spin and thus anyon statistics when the model is coupled to the Chern-Simons term.

Since the vortex excitations in the $x-y$ model are directly associated with the occurence of the phase transition, in this paper we consider the three-dimensional $x-y$ model coupled to the Chern-Simons term. The question we shall investigate here is how this topological term influences the critical properties of this new model and if they are different from the $x-y$ model itself. In order to introduce the Chern-Simons term in the three-dimensional $x-y$ model we need to identify a current. The natural choice in this case is the vortex current which, since it is represented by the vortex lines, can be easily identified. Since the vortex current is conserved we can write it as the curl of a vector potential $\vec{A}$ which enters into the definition of the Chern-Simons term. It turns out that each single vortex line gives a contribution to the Chern-Simons term which is the product of the vortex charge and the winding number of the internal phase angle (cf. the next section) along that vortex line. This interpretation of the Chern-Simons term enables us to compute this term for an arbitrary lattice field configuration. We then couple the Chern-Simons term to the $x-y$ model and compute the helicity modulus and the specific heat for this new model. The finite-size analysis of these quantities leads us to the conclusion that the $x-y$ model with the Chern-Simons term exhibits the same critical properties as the $x-y$ model itself.

The details concerning the introductory remarks above are described in the following sections. In the next section we introduce the three-dimensional $x-y$ model and derive the expression for the Chern-Simons term for this model. The partition function for the $x-y$ model with the Chern-Simons term is also given. Section 3 describes the physical observables as the helicity modulus and the specific heat and discusses the numerical method we employed in order to compute these quantities. Our results are given in section 4. In the last section we briefly summarize our findings.

## 2 The Chern-Simons term in the $x-y$ model

In this section we define the $x-y$ model and give a geometrical interpretation of the Chern-Simons term for this model.

The partition function of the $x-y$ model on the lattice is defined as:

$$Z = \int \prod_j d\theta_j \ \exp(-\beta\mathcal{H}), \tag{1}$$

where $\beta = J/k_B T$ and

$$\mathcal{H} = -\sum_{\langle i,j \rangle} \vec{s}_i \cdot \vec{s}_j . \tag{2}$$

The sum in (2) runs over nearest neighbors, $\vec{s}_i = (\cos\theta_i, \sin\theta_i)$ is a two-component vector which is constrained to be on the unit circle and $J$ sets the energy scale.

Before we turn to the construction of the Chern-Simons term for the $x-y$ model let us briefly discuss the case of the $O(3)$-model. This model is different from the $x-y$ model only in that the spin space is a unit sphere in three dimensions. The equation of motion of the $O(3)$-model admits soliton solutions also known as skyrmions. Wilczek and Zee [7] showed that the skyrmions can be given arbitrary statistics if the Langrangian of the $O(3)$-model is coupled to the Chern-Simons term. Some of the arguments of Wilczek and Zee can be applied to the case of the $x-y$ model in a straightforward manner. Furthermore we will make use of the "form" language (see for example [9]) because this formalism is very convenient for our purposes.

The Chern-Simons term $H_{CS}$ for an abelian gauge field $A$ is defined as follows:

$$H_{CS} = k \int A \wedge dA, \tag{3}$$

where

$$\begin{aligned} A &= A_\mu dx^\mu, &(4)\\ dA &= \partial_\nu A_\mu dx^\nu \wedge dx^\mu. &(5) \end{aligned}$$

The constant $k$ in (3) will be specified later. The symbols in Eqs. (3)-(5) and in the following are defined as in Ref. [9].

Let us consider a spin configuration with $q$ vortex lines on a $L^3$-lattice with periodic boundary conditions. Each vortex line represents a vortex of an integer vortex charge $n_q \in I$. Due to the periodic boundary conditions all vortex lines are closed. If we travel along a vortex line the spins forming the vortex may rotate in the internal spin space. This vortex line is topologically different from a vortex line, representing the same vortex charge but with a different number of spin rotations. We expect the Chern-Simons number (3) to contain this information.

Given a conserved current $j$ a gauge field $A$ can be constructed by [7]

$$dA = \star j. \tag{6}$$

In our case $j$ is the current 1-form whose support is the vortex lines. We write

$$j = \sum_q j_q, \tag{7}$$

where $j_q$ denotes the $q$-th individual vortex current. The vortex charge $n_q$ is obtained from

$$\int_{\partial C_q} d\theta = 2\pi n_q, \tag{8}$$

$\partial C_q$ is an arbitrary contour enclosing the core of the $q$-th vortex line, $\theta$ is the spin angle. Note also, that

$$\int_{C_q} \star j = 2\pi n_q, \tag{9}$$

where $C_q$ is the surface bounded by the curve $\partial C_q$. We want the gauge field $A$ to satisfy the following condition:

$$\int_{\partial C_q} A = 2\pi n_q, \tag{10}$$

or equivalently by the Gauss-Bonnet theorem

$$\int_{C_q} dA = 2\pi n_q. \tag{11}$$

Let us try the ansatz

$$A = d\theta + \sum_q \omega_q, \tag{12}$$

where the support of the 1-forms $\omega_q$ is also restricted to the vortex lines. Using

$$d\omega_q = \star j_q \tag{13}$$

we are able to satisfy (10) and (11).

Eq. (3) takes the following form now:

$$\begin{aligned} H_{CS} &= k \int \left( d\theta + \sum_q \omega_q \right) \wedge \sum_{q'} d\omega_{q'} && (14) \\ &= k \sum_q \int d\theta \wedge d\omega_q + k \sum_q \int \omega_q \wedge d\omega_q . && (15) \end{aligned}$$

Let us spread out the vortex lines, i.e. we define the support of the forms $\omega_q$ to be a thin tube of radius $\epsilon \to 0$ around the vortex lines. Introducing a co-moving 3-bein as a local coordinate system to express $\omega_q$ locally, we realize that $\omega_q$ has components only in a disc perpendicular to its vortex line, or in other words, the last term in (15) vanishes. We are left with

$$H_{CS} = k \sum_q \int d\theta \wedge d\omega_q \tag{16}$$

or, according to the Poincare duality, with

$$H_{CS} = k \sum_q 2\pi n_q \int_{l_q} d\theta, \tag{17}$$

where we integrate along the $q$-th vortex line now. Since $\theta$ is not defined at the center of a vortex we have to give the integral

$$\int_{l_q} d\theta \tag{18}$$

a well defined meaning. Namely, instead of integrating along the actual vortex line we integrate along a line infinitesimally shifted from the original vortex line. The integral (18) measures the spin rotations $m_q$, and we obtain:

$$H_{CS} = k 4\pi^2 \sum_q n_q m_q . \tag{19}$$

If we choose the constant $k$ in (3) to be $1/4\pi^2$, we can define the Chern-Simons number $N_{CS}$ as:

$$N_{CS} = \sum_q n_q m_q . \tag{20}$$

From here we can immediately see that $N_{CS}$ changes its sign under parity or time-inversal transformations. Let us consider the $z$-axis as the euclidean time axis. Then we can think of our model as describing vortices moving in time. A closed loop corresponds to the creation of a vortex and antivortex which move apart and are annihilated after some time. Time-inversion inverses the path in the integral (18), i.e. changes the sign of $m_q$, but leaves the vortex charge $n_q$ unchanged, whereas a parity transformation inverses the path in the integral (10), i.e. changes the sign of $n_q$, but leaves the spin rotations $m_q$ unchanged.

In order to determine the Chern-Simons number $N_{CS}$ for a configuration of spins $\{\vec{s}_i\}$ on the lattice we need to measure the spin rotations $m_q$ along the vortex lines. The vortex lines are perpendicular to the plaquettes and can be found by making use of the conservation of the vortex current, if a cube has an ingoing vortex current, it has an outgoing vortex current as well. The spin rotations $m_q$ can be measured as follows. After introducing a co-moving 3-bein which moves along the vortex line and fixing the position of an arbitrary spin with respect to the 3-bein, we determine how often a spin at our fixed position rotates when we travel along the vortex line.

In the presence of the Chern-Simons term $H_{CS}$ we define the partition function of the $x-y$ model as follows [7]:

$$Z = \int \prod_j d\theta_j \ \exp(-\beta\mathcal{H} + i\alpha N_{CS}). \tag{21}$$

Since we expect the probabilities to find spin configurations with $+N_{CS}$ and $-N_{CS}$ to be equal we restrict the angle variable $\alpha$ to the interval $\alpha \in [0, \pi]$. Let us now consider a vortex with charge $\pm 1$. If we rotate this vortex adiabatically through $2\pi$ over a long period of time the contribution to the partition function (21) is $\exp(i\alpha)$, i.e. the vortex has an internal angular momentum $\alpha/2\pi$. For $\alpha = \pi$ this vortex behaves like a fermion. Since we identify a vortex on a plaquette by computing the winding number of the phase angle $\theta$ around that plaquette according to equation (8) the vortices have only charges $\pm 1$; we can give these vortices an arbitrary internal angular momentum or spin depending on the value of $\alpha$. If we again think of our model as describing vortices moving in euclidean time and if we only consider vortices of charge $\pm 1$, it seems tempting to assume that now the lowest energy state consists of pairs of vortices with opposite charge and opposite spin. This pairing mechanism has its origin in the interaction between the vortices and the additional interaction between the spins. Thus an attractive potential due to the opposite spins in addition to the attractive potential due to the opposite vortex charges between the two paired vortices is introduced. As an intuitive example we can consider two particles of opposite charge moving in a plane. If the particles have spins $\vec{\mu}_1$ and $\vec{\mu}_2$, respectively, in addition to the logarithmic attraction due to the opposite charges the particles feel the interaction energy $V(r) = f(r)\vec{\mu}_1 \cdot \vec{\mu}_2$, where $r$ denotes the distance between the particles and $f(r)$ the $r$-dependent coupling constant between the spins. We expect $f(r) > 0$ which favors pairing of opposite spins. In order to separate such a pair to infinite distance, we need to overcome the attraction due to the opposite charges and the attraction due to the spin–spin interaction. Thus the separation energy for the particles with opposite spins is higher than for the particles without spins. Thus the critical temperature where infinitely separated vortex pairs break up increases with the increasing value of the spin. Since $|\vec{\mu}| \propto \alpha$ and thus $V \propto \alpha^2$ the critical temperature $T_c(\alpha)$ should grow as:

$$T_c(\alpha) = T_c(0) + g\left(\frac{\alpha}{\pi}\right)^2, \tag{22}$$

where $g$ is a constant and $T_c(0)$ is the critical temperature of the three-dimensional $x-y$ model. This is what we indeed find in our numerical study of the model (21) as described in section 4.

# 3 The physical Observables and Monte-Carlo method

The physical quantities we would like to compute for the model (21) are the specific heat $c$ and the helicity modulus $\Upsilon_\mu$. The specific heat is obtained by

$$c = \beta^2 \frac{\langle \mathcal{H}^2 \rangle - \langle \mathcal{H} \rangle^2}{V}, \tag{23}$$

where $V$ denotes the volume of the lattice. The helicity modulus is defined as follows [10, 11]:

$$\frac{\Upsilon_\mu}{J} = \frac{1}{V}\left\langle \sum_{\langle i,j \rangle} \cos(\theta_i - \theta_j)(\vec{e}_\mu \cdot \vec{\epsilon}_{ij})^2 \right\rangle$$

$$-\frac{\beta}{V}\left\langle\left(\sum_{\langle i,j\rangle}\sin(\theta_i-\theta_j)\vec{e}_\mu\cdot\vec{\epsilon}_{ij}\right)^2\right\rangle, \tag{24}$$

$\vec{e}_\mu$ is the unit vector in the corresponding bond direction and $\vec{\epsilon}_{ij}$ is the vector connecting the lattice sites $i$ and $j$. Since our system is isotropic we will omit the vector notation for the helicity modulus in the following.

The expectation values in (23) and (24) are computed with respect to the partition function $Z$ given by the expression (21), i.e. the expectation value of a physical observable $O$ is obtained according to:

$$\langle O\rangle = Z^{-1}\int\prod_j d\theta_j\ O[\theta]\exp(-\beta\mathcal{H}+i\alpha N_{CS}), \tag{25}$$

In order to compute expectation values (25) using the Monte-Carlo method we proceed as in [12]. The partition function given by Eq. (21) can be rewritten as follows:

$$Z=\sum_{N_{CS}}\rho(N_{CS})\exp(i\alpha N_{CS}), \tag{26}$$

with $\rho(N_{CS})$ being the density of the configurations whose value for the Chern-Simons number is $N_{CS}$. This density is given by:

$$\rho(N_{CS})=\int\prod_i d\theta_i\Big|_{N_{CS}}\exp(-\beta\mathcal{H}). \tag{27}$$

The integration on the right hand side of Eq. (27) includes only those configurations whose value for the Chern-Simons number is $N_{CS}$. The expectation value (25) can then be expressed as

$$\langle O\rangle = Z^{-1}\sum_{N_{CS}}\rho(N_{CS})\langle O\rangle_{N_{CS}}\exp(i\alpha N_{CS}), \tag{28}$$

where

$$\langle O\rangle_{N_{CS}}=\frac{1}{\rho(N_{CS})}\int\prod_i d\theta_i\Big|_{N_{CS}}\exp(-\beta\mathcal{H}). \tag{29}$$

In a real computer simulation we are able to compute $\rho(N_{CS})$, $Z$ and $\langle O\rangle_{N_{CS}}$ to a sufficient accuracy. Therefore using (28) we can compute the average value $\langle O\rangle$. For our simulation we assume $\rho(N_{CS})$ and $\langle O\rangle_{N_{CS}}$ to be even functions with respect to $N_{CS}$, i.e. all expectation values are real.

The simulation is done as follows. We generate a spin configuration according to the probability distribution

$$P[\theta]\propto\exp(-\beta\mathcal{H}), \tag{30}$$

by using Wolff's 1-cluster algorithm [13]. We measure the Chern-Simons number $N_{CS}$ of this configuration, increment $\rho(N_{CS})$ by 1 and add the value of the observable $O(N_{CS})$ to a table containing all the values $O(N_{CS})$ labeled by the Chern-Simons numbers. Next a new spin configuration is created. The histogram $\rho(N_{CS})$ is proportional to the distribution (27), the arithmetic average of the values $O(N_{CS})$ for a given value of $N_{CS}$ yields an estimate for $\langle O\rangle_{N_{CS}}$. This allows us to calculate the average value $\langle O\rangle$ given by (28). The accuracy of this method can be improved by using reweighting techniques [14].

We computed the helicity modulus and the specific heat on lattices of sizes $L\times L\times L$ with $L=4,5,6,8,10$. Periodic boundary conditions were applied. We carried out of the order of 50,000 thermalization steps and of the order of 2,000,000 measurements. The computations were performed on a heterogeneous environment of workstations which include Sun, IBM RS/6000 and DEC alpha AXP workstations. Due to the reweighting procedure and the broadening of the densitiy distribution $\rho(N_{CS})$ with larger lattices, runs on lattices with $L\gg 10$ become extremely time consuming. Therefore we restricted ourselves to the above lattice sizes.

# 4 Results

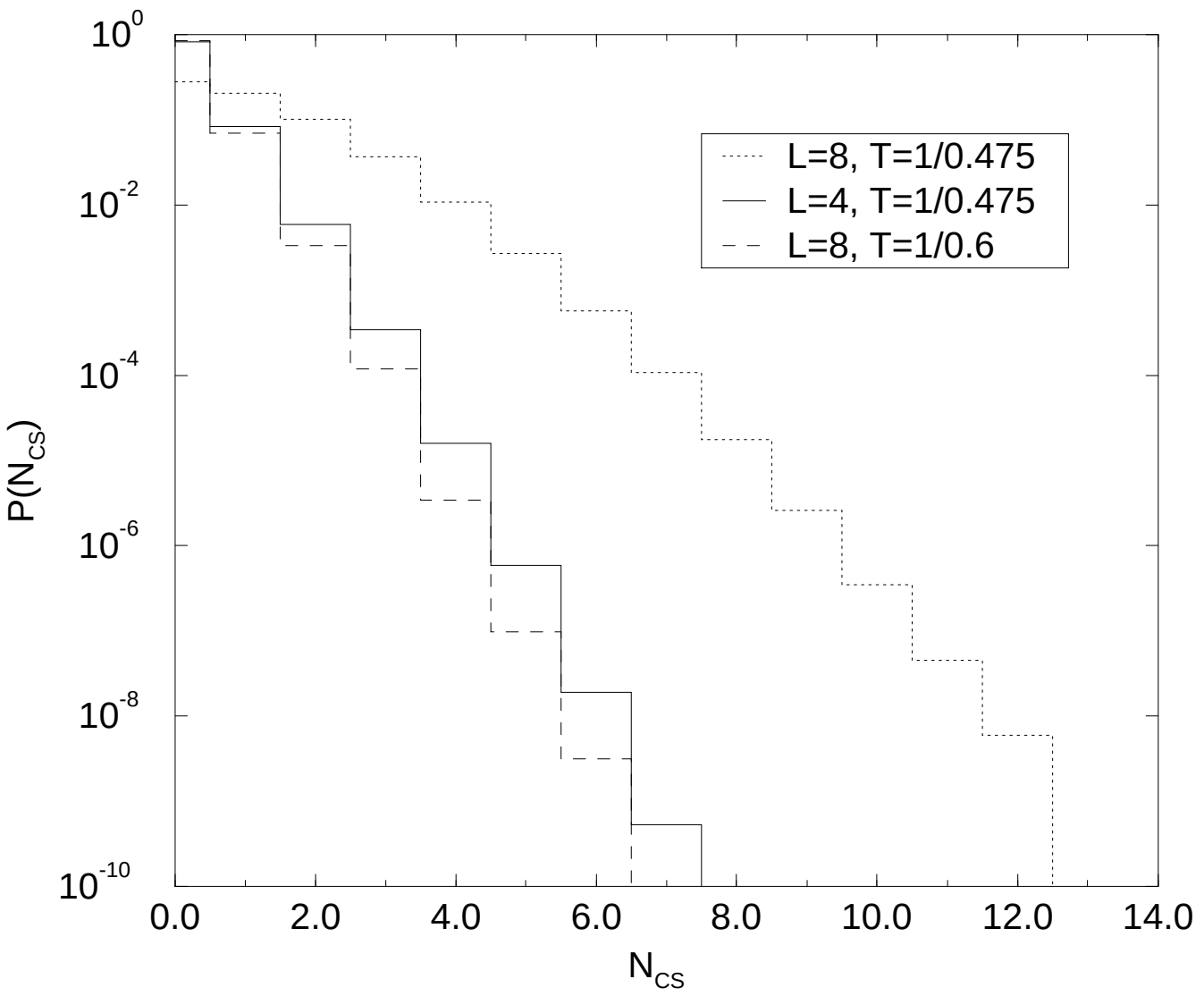

Figure 1: The probability distribution $P(N_{CS})$ for different lattice sizes and temperatures.

Let us first calculate the normalized probability distribution of the Chern-Simons number $P(N_{CS})$ defined as:

$$P(N_{CS}) = \frac{\rho(N_{CS})}{\sum_{N_{CS}} \rho(N_{CS})}, \tag{31}$$

where $N_{CS}$ and $\rho(N_{CS})$ are defined by Eqs. (20) and (27). Fig.1 shows the distribution function $P(N_{CS})$ for different lattice sizes and different temperatures. As can be seen $P(N_{CS})$ is strongly affected by the size of the system and the temperature. Namely, by increasing the lattice size or the temperature we observe a broadening of the probability distribution $P(N_{CS})$.

In order to create the Chern-Simons number $N_{CS}$ a certain amount of energy $E_{N_{CS}}$ is required. If we assume this energy to be independent of $N_{CS}$ we can write for $P(N_{CS})$:

$$P(N_{CS}) \propto \exp(-\beta E_{N_{CS}} N_{CS}). \tag{32}$$

Thus the slope $N_{CS}^{-1} \log P(N_{CS}) = -\beta E_{N_{CS}}$. The constant slopes can be seen in Fig.1 for large values of $N_{CS}$. The energy $E_{N_{CS}}$ itself depends on the lattice size $L$ and the temperature.

## 4.1 The helicity modulus

Here we investigate the finite-size scaling behavior of the helicity modulus for different values of the parameter $\alpha$. We use the finite-size scaling properties of the helicity modulus to extract an estimate for the $\alpha$-dependent critical temperature $T_c(\alpha)$ and to study the $\alpha$-dependence of the critical exponent $\nu$ of the helicity modulus.

At this point we would like to mention one difficulty we encountered in the course of our computations. Because of the imaginary part in the partition function (21) the error bars of the computed quantities grow with increasing values of $\alpha$. As long as the distribution $P(N_{CS})$ is sharply peaked around $N_{CS} = 0$ the error bars are reasonably small. This happens for small temperatures and small lattice sizes (cf. Fig.1 for the different shapes of $P(N_{CS})$ and Fig.2 and Fig.3 for the size of the error bars). For larger temperatures and larger lattices $P(N_{CS})$ grows broader and broader which requires longer and longer simulation times in order to keep the error bars reasonably small. Therefore we restricted our computer simulations to lattices with at most 1000 lattice sites.

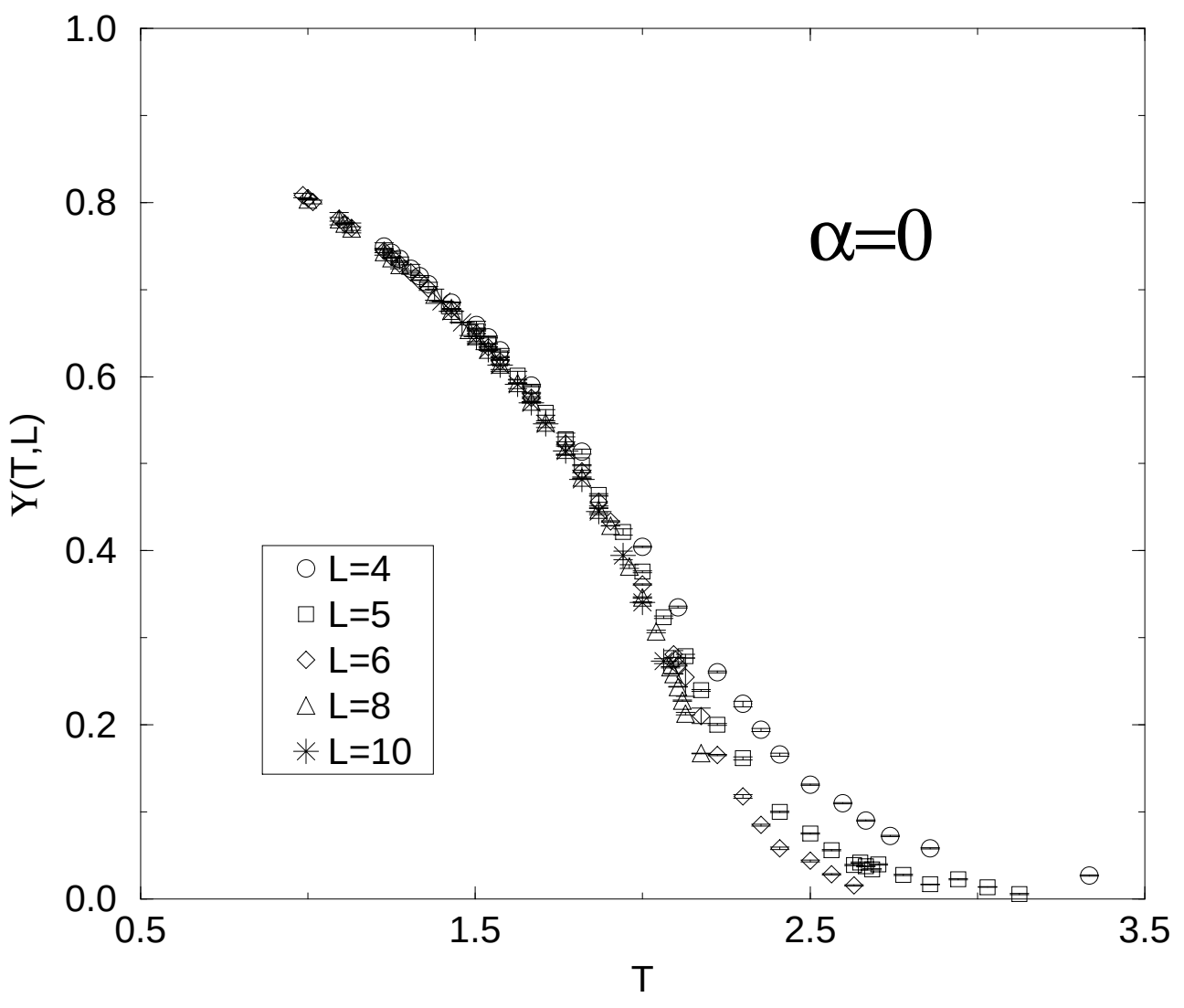

Figure 2: The helicity modulus $\Upsilon(T,L)$ as a function of $T$, $\alpha = 0$.

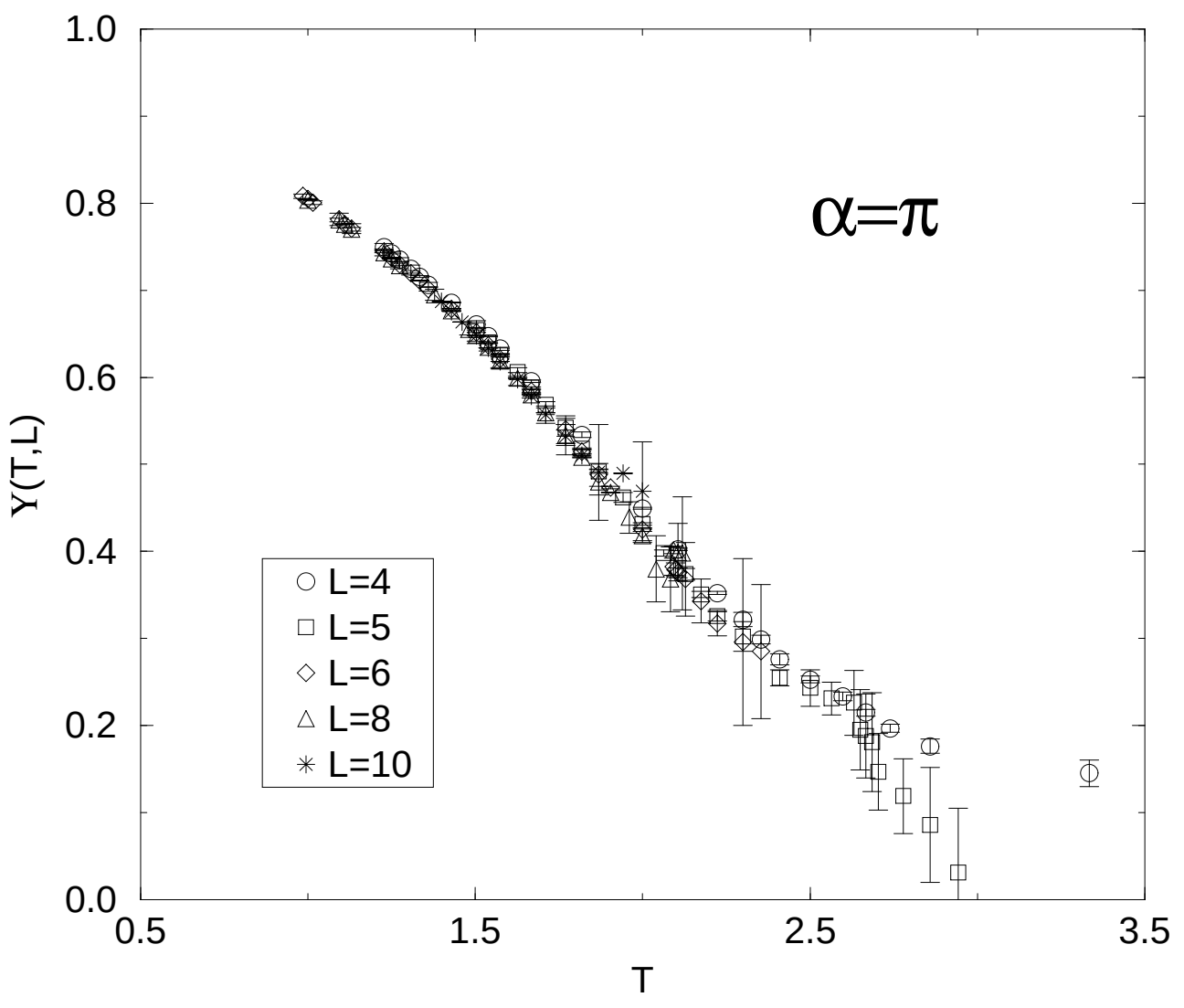

Figure 3: The helicity modulus $\Upsilon(T,L)$ as a function of $T$, $\alpha = \pi$.

In Fig.2 and Fig.3 we plot our data for the helicity modulus with respect to temperature for $\alpha = 0$ and $\alpha = \pi$, respectively. We see that the values of the helicity modulus corresponding to $\alpha = \pi$ are larger than the ones pertaining to $\alpha = 0$ for temperatures $T \geq 1.7$. In general we find $\Upsilon(T, L, \alpha_1) \geq \Upsilon(T, L, \alpha_2)$ if $\alpha_1 > \alpha_2$. For low temperatures the values of $\Upsilon(T, L)$ for both $\alpha = 0$ and $\alpha = \pi$ are about the same. Thus the spin stiffness becomes larger when the vortices have a non-zero spin. Therefore we expect the critical temperature to grow with increasing $\alpha$.

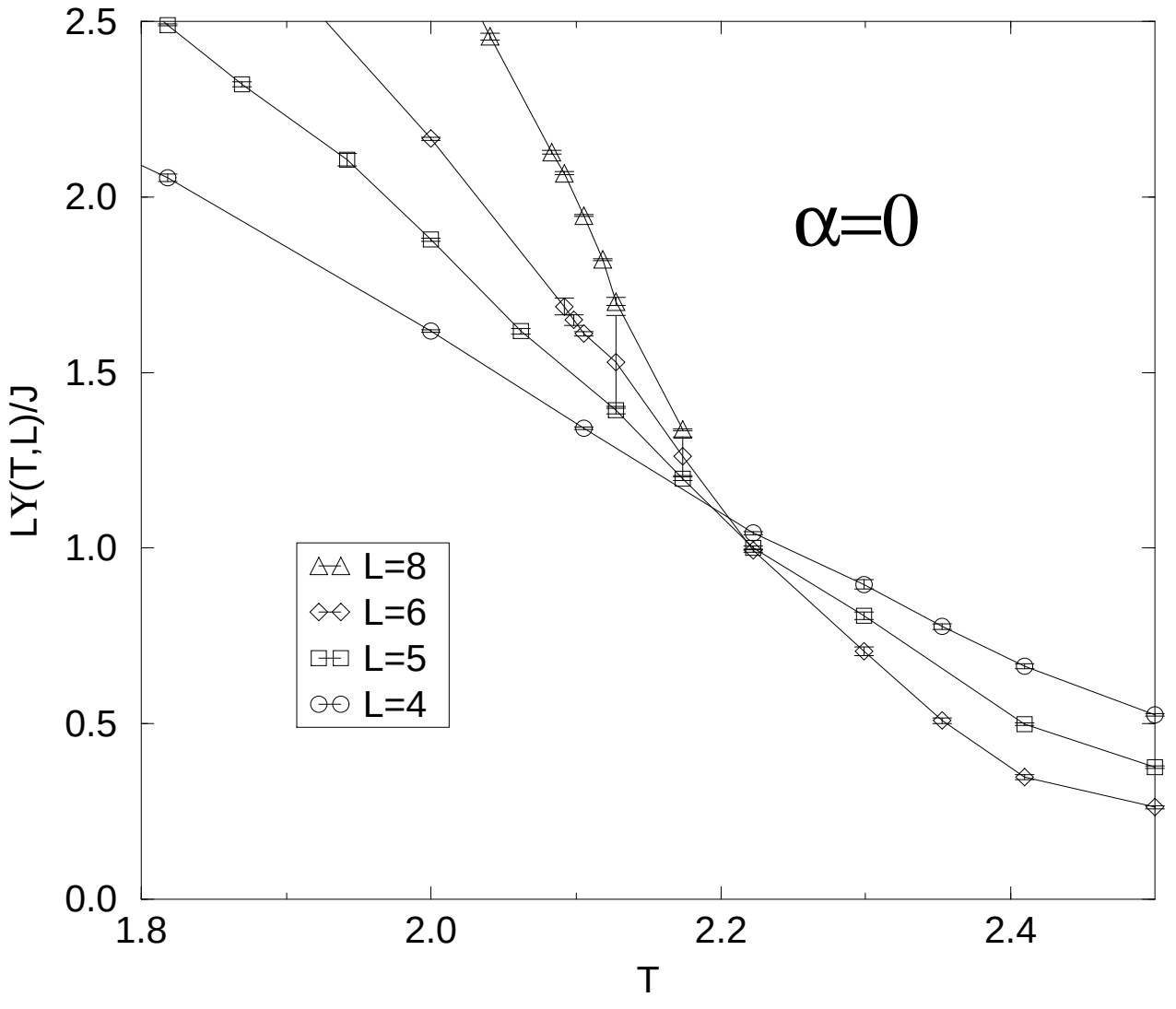

Figure 4: $L\Upsilon(T, L)/J$ as a function of $T$, $\alpha = 0$.

In order to find a rough estimate for the critical temperatures at various $\alpha$ we plot the dimensionless quantity $L\Upsilon(T, L, \alpha)/J$ versus $T$. Following the finite-size scaling theory [15] we expect to find that

$$\frac{L\Upsilon(T, L, \alpha)}{J} = f(\frac{L}{\xi(T)}, \alpha) \tag{33}$$

close to the critical temperature $T_c(\alpha)$. $\xi(T)$ denotes the correlation length of the infinitely extended system and $f$ is a universal function. At $T_c(\alpha)$ the correlation length is infinite, thus for fixed $L$ the right hand side of Eq. (33) is a constant. Therefore all the curves $L\Upsilon(T, L, \alpha)/J$ intersect in one point whose abscissa is $T_c(\alpha)$. In Fig.4 and Fig.5 we show $L\Upsilon(T, L, \alpha)/J$ versus $T$ for $\alpha = 0$ and $\alpha = \pi/2$, respectively. From the plots we read off $T_c(0) = 2.206(12)$ and $T_c(\pi/2) = 2.329(11)$. The errorbars where estimated with respect to the scattering of the intersection points. Using the described method we obtained estimates of the critical temperatures for $\alpha = 0, \pi/4, \pi/2, 3\pi/4$ (cf. Table 1). For $\alpha = \pi$ this method of determining $T_c$ is not accurate enough due to the large error bars of $L\Upsilon(T,L)/J$. We will determine $T_c(\pi)$ approximately using a different technique as discussed further below.

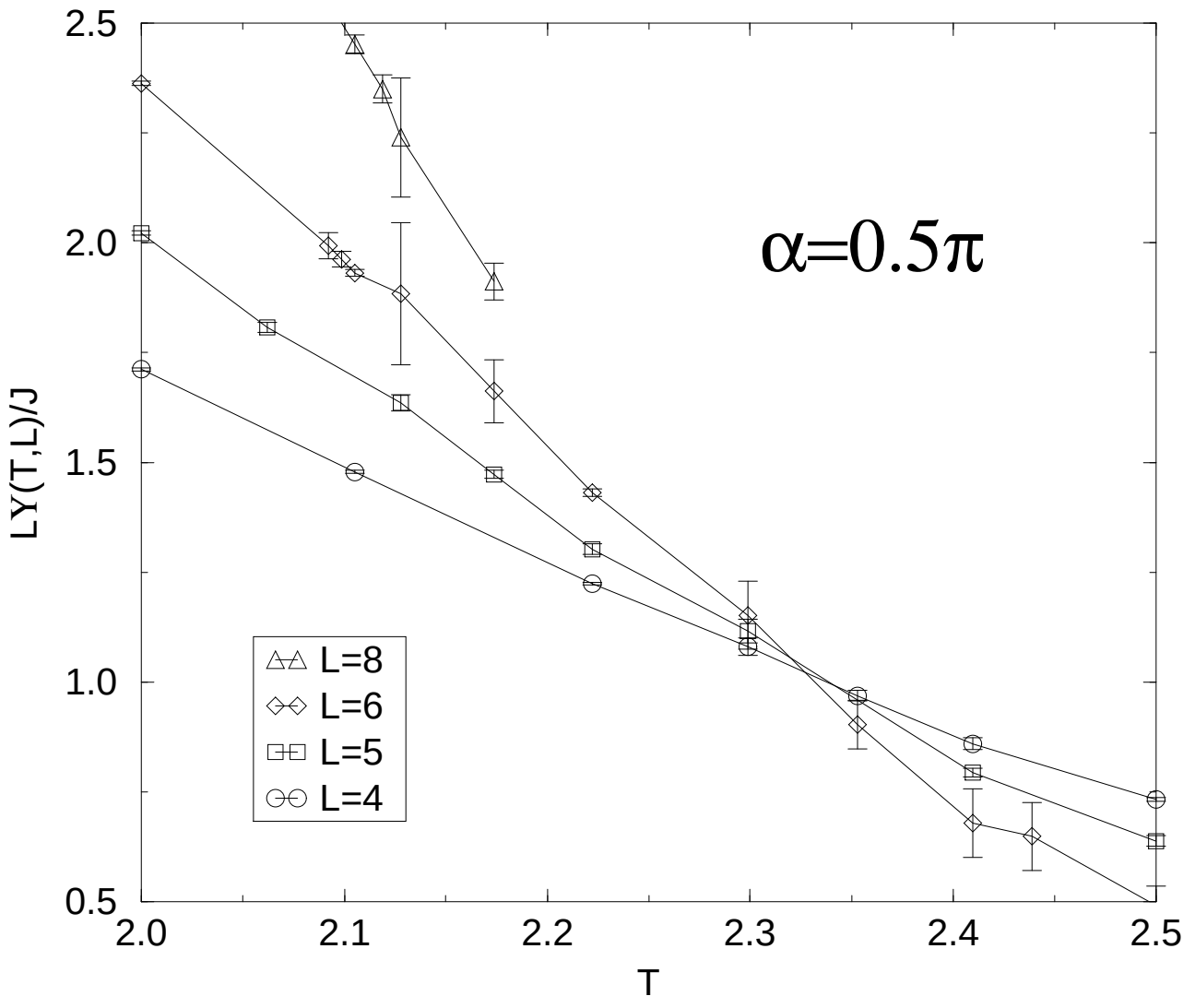

Figure 5: $L\Upsilon(T,L)/J$ as a function of $T$, $\alpha = \pi/2$.

| $\alpha$ | $T_c(\alpha)$ |
|---|---|
| 0 | 2.206(12) |
| $\frac{\pi}{4}$ | 2.227(13) |
| $\frac{\pi}{2}$ | 2.329(11) |
| $\frac{3\pi}{4}$ | 2.52(8) |
| $\pi$ | 2.65(10) |

Table 1: The critical temperature for different values of $\alpha$.

In order to check if the critical exponent $\nu$ becomes $\alpha$-dependent let us explore formula (33) a little more. From the finite-size scaling theory we know that [15]

$$\xi(t \to 0) = A(\alpha)|t|^{-\nu(\alpha)}, \tag{34}$$

where the reduced temperature $t = 1 - T/T_c(\alpha)$. We introduced a possible $\alpha$-dependence of the constant $A$ and the critical exponent $\nu$ in formula (34). We have $\nu(0) = 0.6705$ [16, 17, 18]. With the help of (34) we may rewrite expression (33) as follows:

$$\frac{L\Upsilon(T,L,\alpha)}{J} = \tilde{f}(tL^{1/\nu(\alpha)}, \alpha). \tag{35}$$

Eq. (35) means that we obtain one universal curve for all values of $L$ if we plot $L\Upsilon(T,L,\alpha)/J$ versus $tL^{1/\nu(\alpha)}$. This is indeed the case for $\alpha = 0$ which is demonstrated in Fig.6. The data collapse on one universal curve for $tL^{1/\nu(0)} > -1.6$, for the plot we used $T_c(0) = 2.206$ (cf. Table 1) and $\nu(0) = 0.6705$. We are now in a position to check the $\alpha$-dependence of the critical exponent $\nu$. If the inequility $\nu(\alpha) \neq 0.6705$ is true we will not find a universal curve by plotting $L\Upsilon(T,L,\alpha)/J$ versus $tL^{1/0.6705}$. In Fig.7 we demonstrate the collapse of our data points for $\alpha = \pi/2$ with $T_c(\pi/2) = 2.329$ (cf. Table 1) and $\nu(\pi/2) = 0.6705$. We found this collapse of our data also for $\alpha = \pi/4$ and $\alpha = 3\pi/4$ when the value of $\nu = 0.6705$ was taken. Therefore we conclude that $\nu$ is not affected when the $x-y$ model is coupled to the Chern-Simons term defined in Section 2 according to (21). This enables us to estimate the critical temperature $T_c(\pi)$, namely, we plot $L\Upsilon(T,L)/J$ versus $tL^{1/0.6705}$ but vary the value for $T_c(\pi)$ until we are able to collapse our data points on one universal curve. This can be achieved reasonably well for $T_c(\pi) = 2.65(10)$. The scaling function thus obtained is displayed in Fig.8. We notice that the range of $tL^{1/0.6705}$ where the data collapse on one universal curve seems larger for larger $\alpha$.

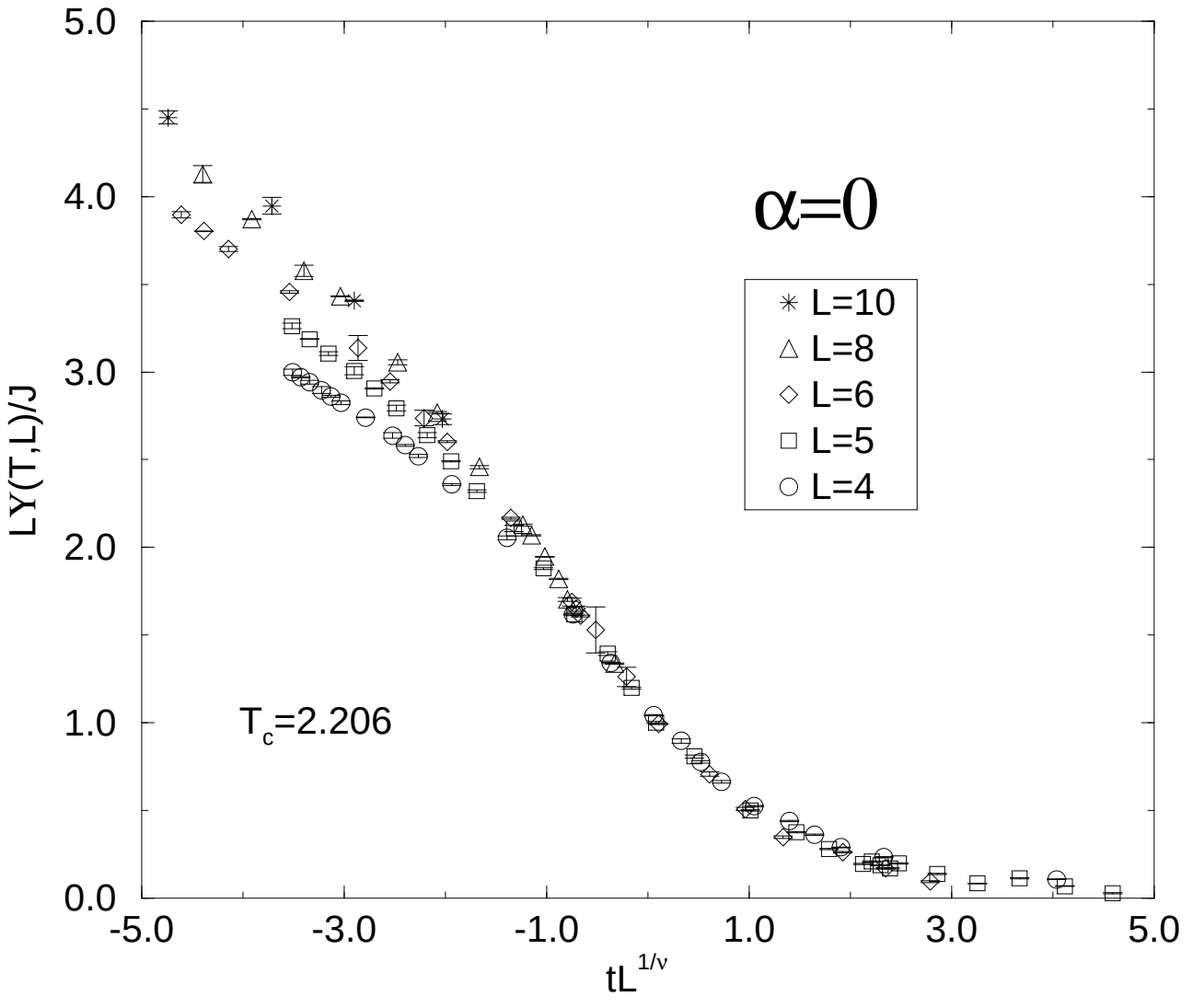

Figure 6: $L\Upsilon(T,L)/J$ as a function of $tL^{1/\nu}$ for $\alpha = 0$, $\nu = 0.6705$, $T_c = 2.206$.

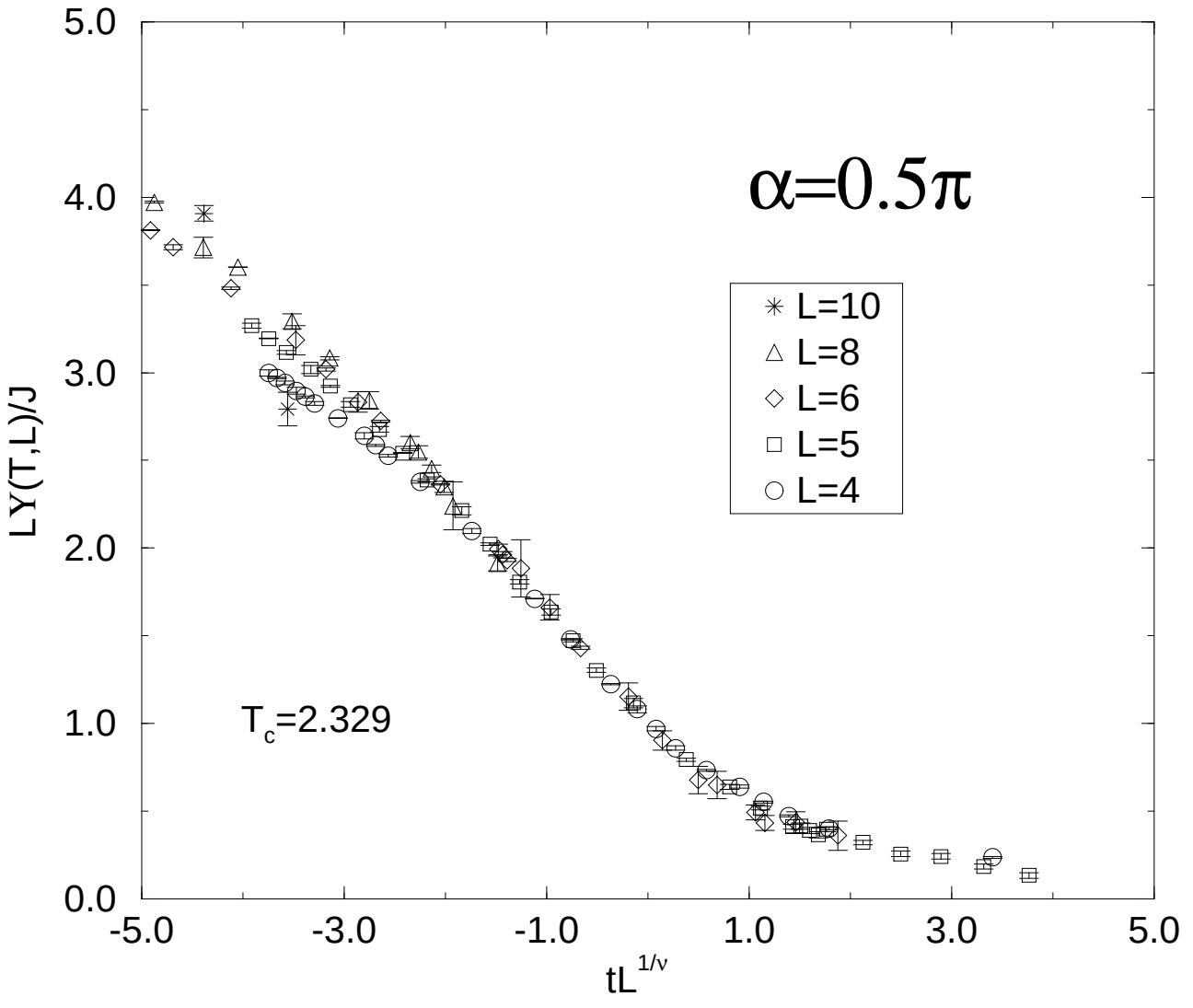

Figure 7: $L\Upsilon(T,L)/J$ as a function of $tL^{1/\nu}$ for $\alpha = \pi/2$, $\nu = 0.6705$, $T_c = 2.329$.

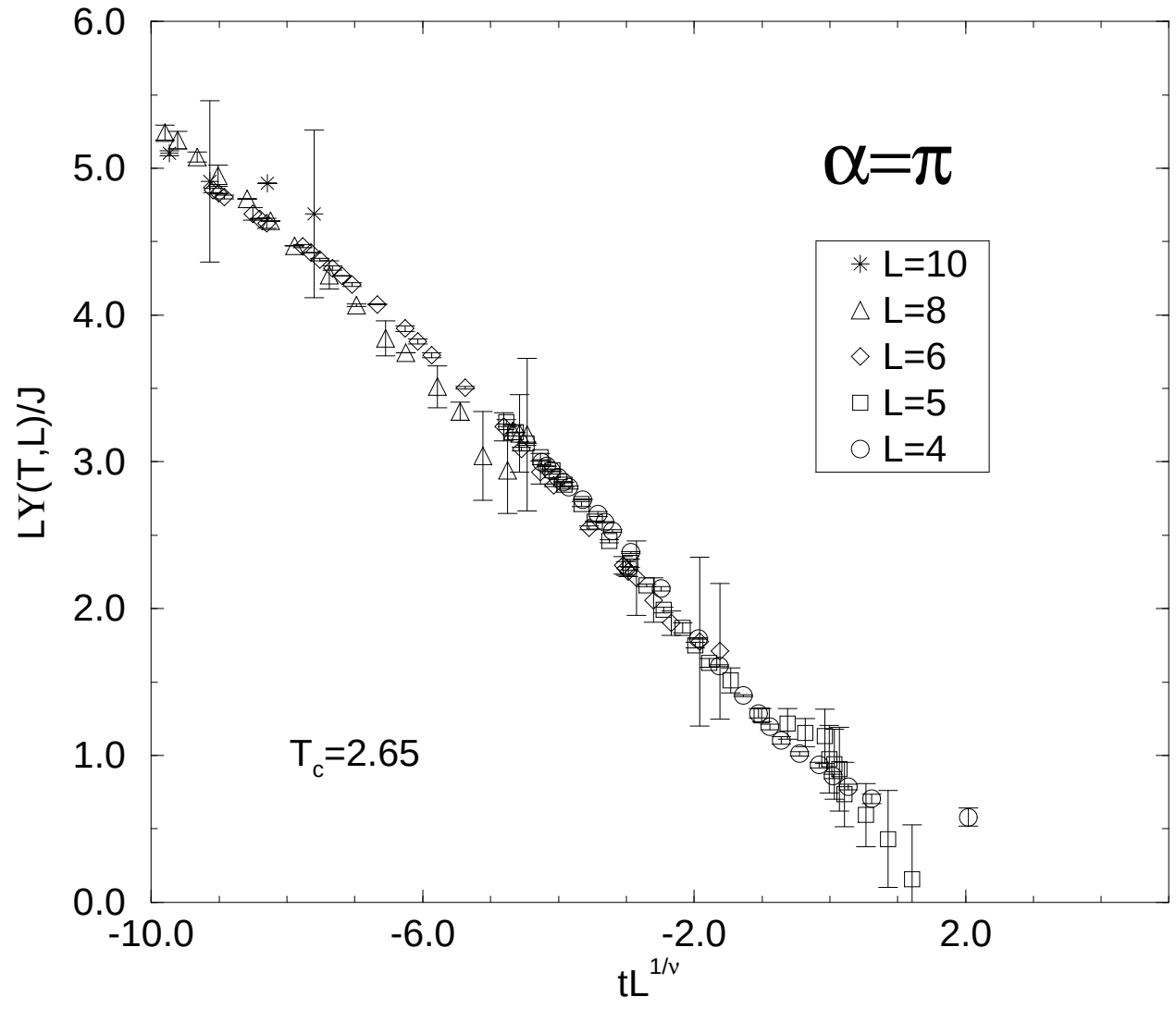

Figure 8: $L\Upsilon(T,L)/J$ as a function of $tL^{1/\nu}$ for $\alpha=\pi$, $\nu=0.6705$, $T_c=2.65$.

In Fig.9 we notice that the critical temperature $T_c(\alpha)$ increases monotonically with increasing $\alpha$. In order to check whether our values for $T_c(\alpha)$ are consistent with the expression (22) we fitted the functional form (22) to the critical temperature values given in Table 1 with the fixed parameter $T_c(0) = 2.206$. We obtained $g = 0.484 \pm 0.038$ and the fit is given by the parabola in Fig.9. Since the parabola fits the computed values for $T_c(\alpha)$ rather well, we can think of the $x-y$ model with the Chern-Simons term (20) as describing vortices with spin $\pm\alpha/(2\pi)$ moving in euclidean time, which interact through the potential due to the vortex charges and the spin-spin interaction potential $\propto \alpha^2$.

## 4.2 The specific heat

In this section we briefly comment on the behavior of the specific heat with respect to the spin coefficient $\alpha$.

Since the critical exponent of the correlation length $\nu$ does not depend on $\alpha$ (cf. the previous section) the critical exponent of the specific heat $\tilde{\alpha}$ cannot depend on $\alpha$ either due to the hyperscaling assumption $\tilde{\alpha} = 2 - 3\nu$. We denote the critical exponent of the specific heat by $\tilde{\alpha}$ instead of $\alpha$ (which is commonly used in the literature to denote the specific heat critical exponent) in order to avoid confusion with the parameter $\alpha$ of this paper (cf. Eq. (21).

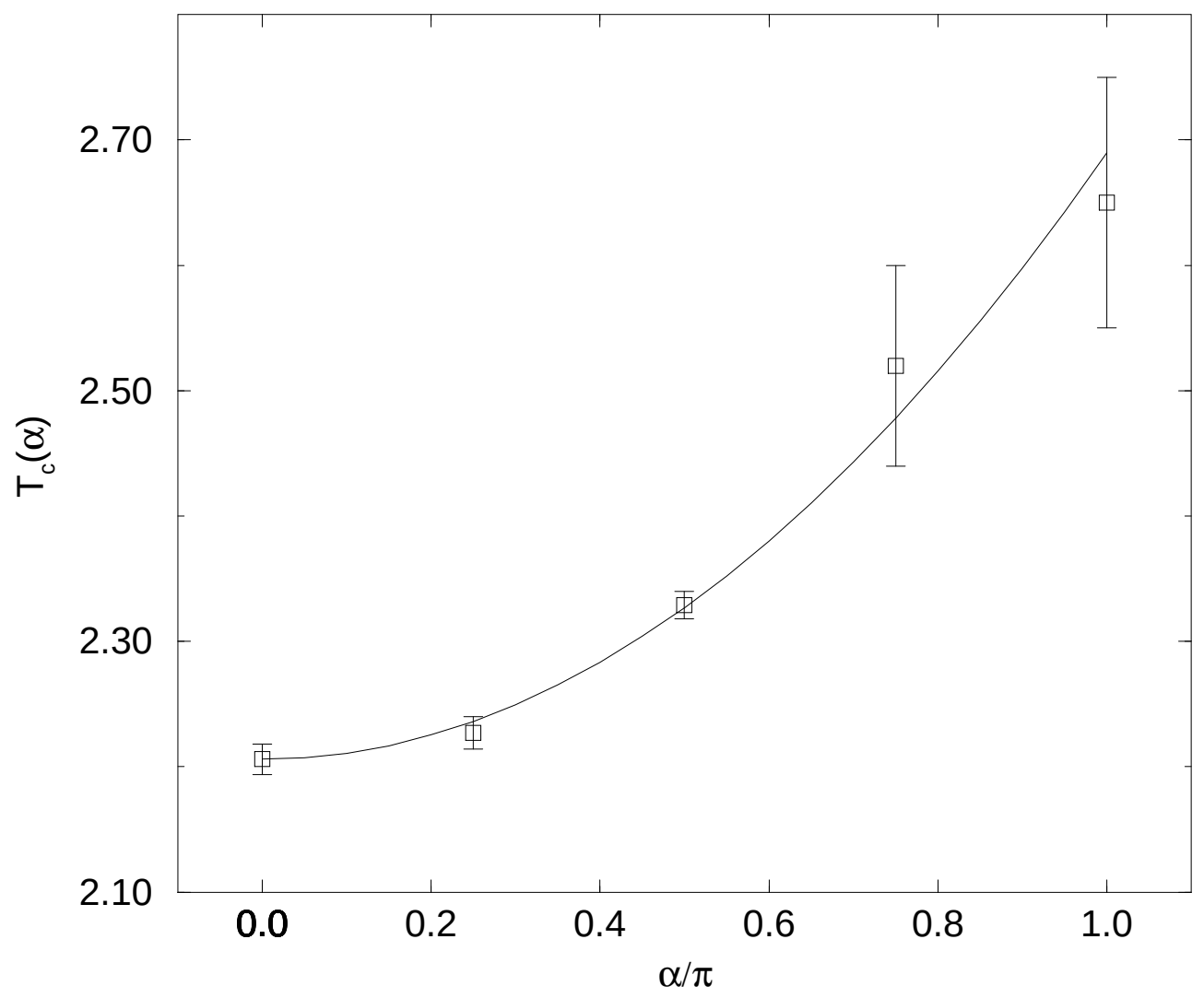

Figure 9: The critical temperatures $T_c(\alpha)$ as a function of $\alpha$. The solid line represents Eq. (22) with $T_c(0) = 2.206$ and $g = 0.484$.

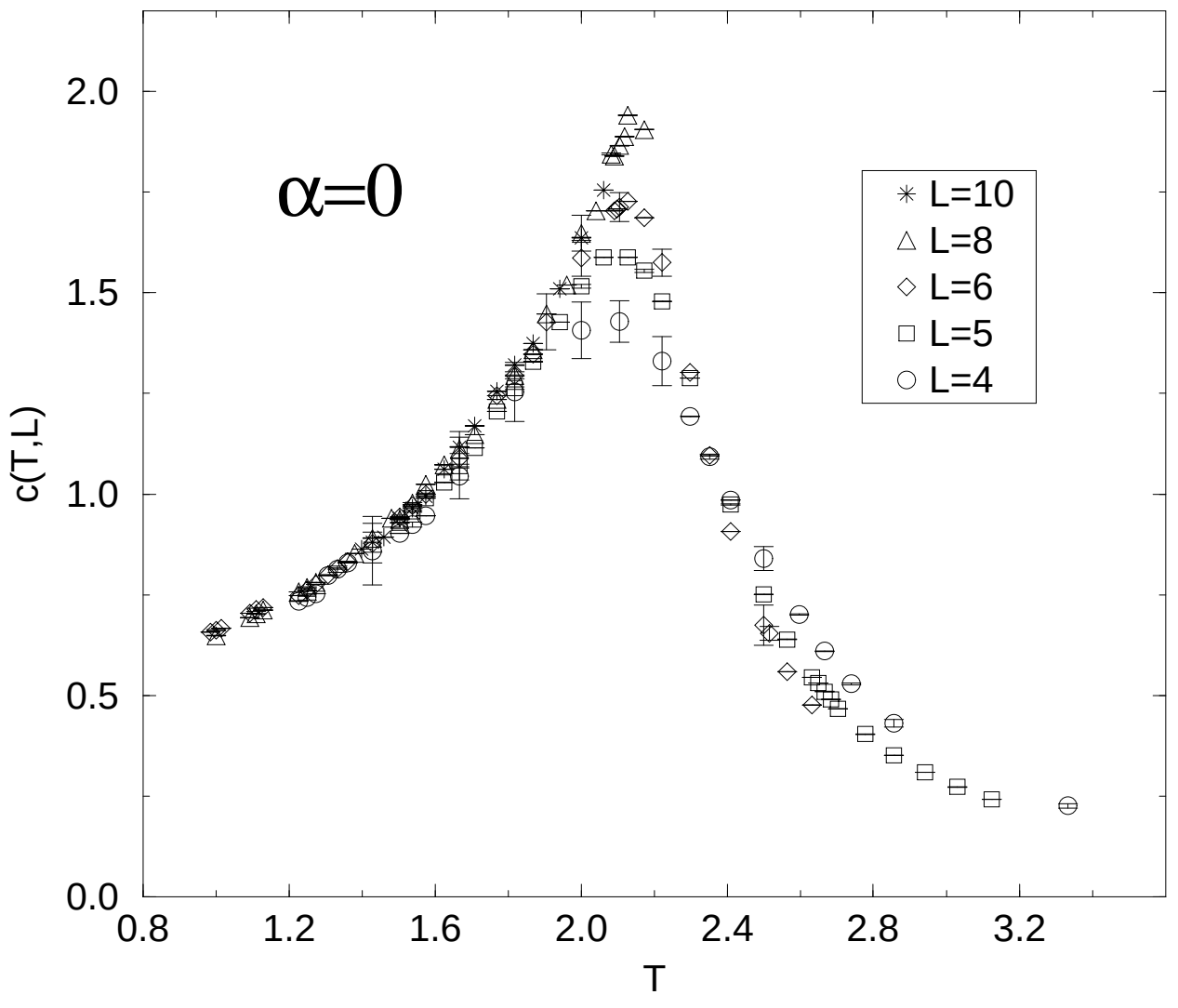

Figure 10: The specific heat for various lattice sizes as a function of temperature at $\alpha = 0$.

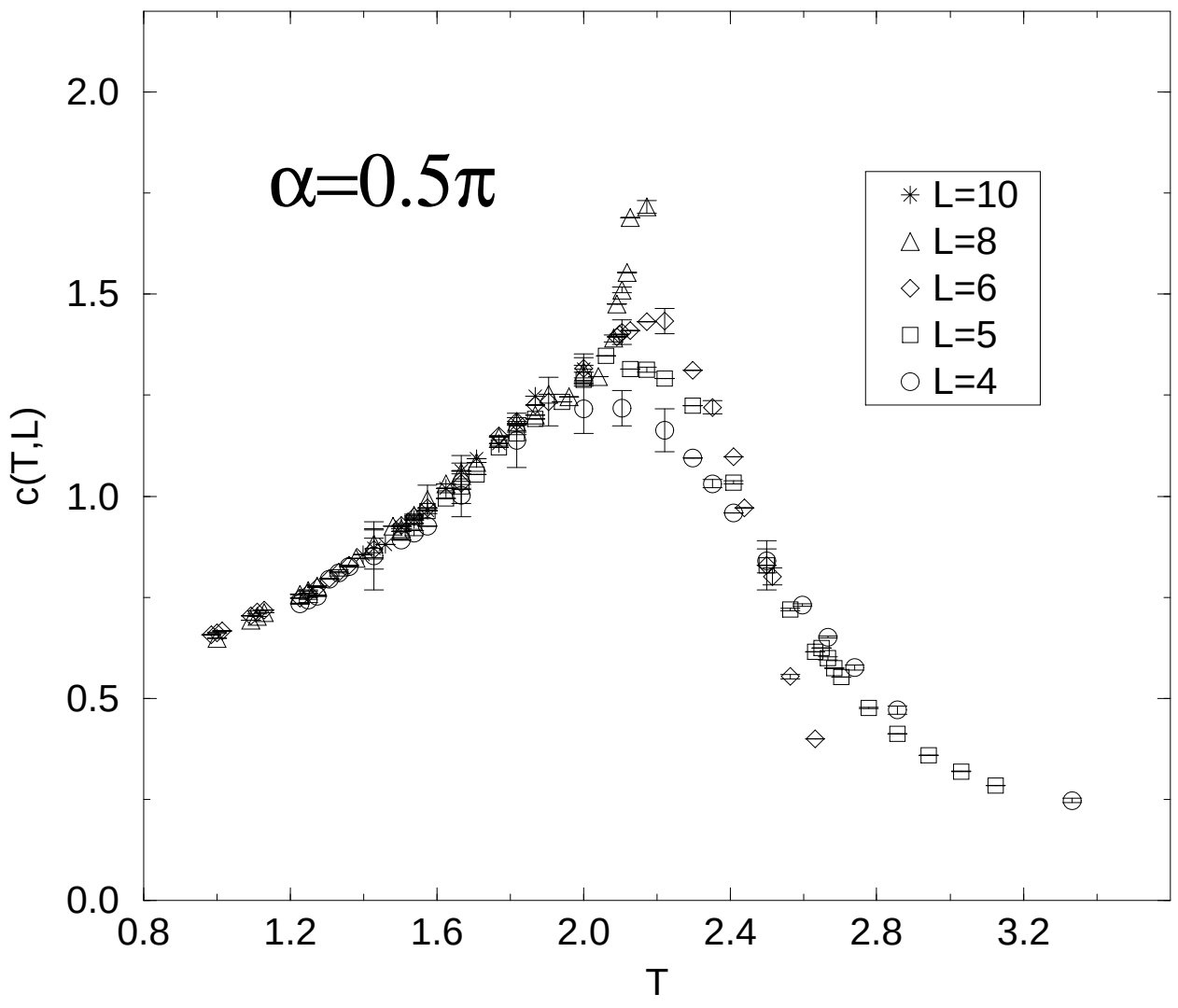

Figure 11: The specific heat for various lattice sizes as a function of temperature at $\alpha = \pi/2$.

Fig.10 and Fig.11 show the specific heat for various lattice sizes at $\alpha = 0$ and $\alpha = \pi/2$, respectively. At low temperatures the specific heat data agree for all lattice sizes and all values of $\alpha$. This should be the case since at these temperatures the correlation length is smaller than the lattice size $L$, thus the specific heat does not feel the finite size of the system. Furthermore the probability of finding the Chern-Simons number $N_{CS} \neq 0$ is very small, i.e. the parameter $\alpha$ does not have any influence on the physical quantities at all. However, for temperatures closer to the critical temperature $T_c(0)$ we find that the values of the specific heat become smaller with increasing $\alpha$. This effect is independent of the system size $L$ (cf. Fig.10 and Fig.11). Our findings seem to indicate that the finite value of the specific heat $c(T_c(\alpha), L = \infty)$ is $\alpha$-dependent and decreases with increasing $\alpha$. However, in order to determine these values to a sufficient accuracy, computations on much larger lattices have to be performed which require very long computation times due to the broadening of the probability distribution $P(N_{CS})$.

# 5 Summary

We have coupled the three-dimensional $x - y$ model to the Chern-Simons term in the spirit of Wilczek and Zee [7], thus endowing the vortices with a spin $\alpha/(2\pi)$. We have investigated the influence of the Chern-Simons term on the critical behavior. The geometrical interpretation of this term enabled us to compute the Chern-Simons number for an arbitrary spin configuration on the lattice. We computed the helicity modulus and the specific heat on $L \times L \times L$ lattices up to $L = 10$ by means of a Monte-Carlo simulation. Periodic boundary conditions were applied in all directions. We used the finite-size scaling properties of the helicity modulus to estimate the critical temperatures and to check the influence of the Chern-Simons term on the critical exponent $\nu$. Our findings suggest that the $x - y$ model coupled to the Chern-Simons term belongs to the $x - y$ universality class, however, the critical temperature and the finite bulk value of the specific heat at the critical temperature depend on $\alpha$.

# 6 Acknowledgements

N.S. would like to thank D. W. Sumners and E. Klassen for discussing the form language and the nature of Chern-Simons terms. This work was supported by the National Aeronautics and Space Administration under grant no. NAGW-3326.